\documentclass[10pt,journal]{IEEEtran}

\ifCLASSOPTIONcompsoc
  \usepackage[nocompress]{cite}
\else
  \usepackage{cite}
\fi

\ifCLASSINFOpdf
  \usepackage[pdftex]{graphicx}
  \DeclareGraphicsExtensions{.pdf,.jpg,.png,.eps,.png}
\else
\fi

\usepackage{amsmath}
\usepackage{amsthm}
\usepackage{color}

\usepackage[normalem]{ulem}

\usepackage{algorithm, algorithmic}

\makeatletter
\newcommand\fs@norules{\def\@fs@cfont{\bfseries}\let\@fs@capt\floatc@ruled
  \def\@fs@pre{}%
  \def\@fs@post{}%
  \def\@fs@mid{\kern3pt}%
  \let\@fs@iftopcapt\iftrue}
\makeatother
\floatstyle{norules}
\restylefloat{algorithm}

\ifCLASSOPTIONcompsoc
  \usepackage[caption=false,font=footnotesize,labelfont=sf,textfont=sf]{subfig}
\else
  \usepackage[caption=false,font=footnotesize]{subfig}
\fi

\begin{document}

\title{Aerial Access and Backhaul in mmWave B5G Systems: Performance Dynamics and Optimization}

	\author{Nikita Tafintsev, Dmitri Moltchanov, Mikhail Gerasimenko, Margarita Gapeyenko, Jing Zhu, Shu-ping Yeh, Nageen Himayat, Sergey Andreev, Yevgeni Koucheryavy, and Mikko Valkama\vspace{-0mm}
		\thanks{Nikita Tafintsev, Dmitri Moltchanov, Mikhail Gerasimenko, Margarita Gapeyenko, Sergey Andreev, Yevgeni Koucheryavy, and Mikko Valkama are with Tampere University, Finland.}
		\thanks{Jing Zhu, Shu-ping Yeh, and Nageen Himayat are with Intel Corporation, Santa Clara, CA, USA.}
		\thanks{This work was supported by Intel Corporation, project 5G-FORCE, and Academy of Finland (projects WiFiUS, PRISMA, and RADIANT).}
		\thanks{\textcopyright 2019 IEEE. Personal use of this material is permitted. Permission from IEEE must be obtained for all other uses, in any current or future media, including reprinting/republishing this material for advertising or promotional purposes, creating new collective works, for resale or redistribution to servers or lists, or reuse of any copyrighted component of this work in other works.}
		}

\maketitle

\begin{abstract}

The use of unmanned aerial vehicle (UAV)-based communication in millimeter-wave (mmWave) frequencies to provide on-demand radio access is a promising approach to improve capacity and coverage in beyond-5G (B5G) systems. There are several design aspects to be addressed when optimizing for the deployment of such UAV base stations. As traffic demand of mobile users varies across time and space, dynamic algorithms that correspondingly adjust the UAV locations are essential to maximize performance. In addition to careful tracking of spatio-temporal user/traffic activity, such optimization needs to account for realistic backhaul constraints. In this work, we first review the latest 3GPP activities behind integrated access and backhaul system design, support for UAV base stations, and mmWave radio relaying functionality. We then compare static and mobile UAV-based communication options under practical assumptions on the mmWave system layout, mobility and clusterization of users, antenna array geometry, and dynamic backhauling. We demonstrate that leveraging the UAV mobility to serve moving users may improve the overall system performance even in the presence of backhaul capacity limitations.

\end{abstract}


\section{Introduction}\label{sect:01}


The unmanned aerial vehicles (UAVs) with their unconstrained 3D mobility and autonomous flight capabilities are becoming adopted across various applications. The telecom sector is among those benefiting from active UAV utilization~\cite{bor20195g,hellaoui2018aerial}. Attractive UAV applications for mobile operators include, for example, base station (BS) inspection, wireless connection in disaster-affected regions, and network densification during temporary mass events.


UAVs acting as BS carriers, named in this work UAV-BSs, have recently gained increased interest from the academic and industrial communities~\cite{gapeyenko_jsac_drones}. This is in part to meet the stringent performance requirements related to ubiquitous coverage, for example, during short-lived and spontaneous events in order to strategically densify the network. Here, the use of conventional BSs may lead to sub-optimal radio resource utilization. Hence, an alternative solution to serve some of the users by the UAV-BSs may boost capacity and improve resource efficiency. Particular interest is dedicated to the UAV-BSs equipped with 5G New Radio (NR) capabilities that are able to support a large number of users while satisfying the desired data rate and latency requirements.


To fully benefit from the utilization of UAV-BSs and optimize the system-level performance, careful placement of the UAV-BSs is essential. There is a number of factors that affect the positioning of the UAV-BSs. One of these is backhaul connectivity between the UAV-BSs and the core network, which may impact the overall system performance. While various algorithms for optimized UAV-BS deployment have been proposed in recent literature~\cite{kalantari2016number}, \cite{iellamo2017placement}, there is a lack of comprehensive studies considering all of the important factors under practical modeling assumptions.  


The use of millimeter-wave (mmWave) bands allows for a dramatic increase in the data rates made available to mobile users~\cite{boccardi2013five}. In addition, mmWave transmission requires highly directional transceivers, which significantly lowers radio interference between the nearby communicating nodes, thus enabling more flexible positioning and higher network density. However, along with its inherent benefits, the use of mmWave frequencies poses unique challenges to wireless system design. Due to their significantly shorter coverage, UAV-BSs need to closely track the spatio-temporal traffic peaks across the network. This is related to increased backhaul dynamics and more frequent changes in the effective network topology.


The goal of this work is to evaluate the performance of UAV-aided radio systems enabled by integrated access and backhaul (IAB) capabilities with the aid of system-level simulations.  The IAB technology is employed in terms of mmWave spectrum utilization for both UAV-BS to user equipment (UE) access and UAV-BS to ground cell backhaul connections. Our evaluation concentrates on realistic deployments with moving and clustered users, practical antenna arrays at both the UAV-BS and the UE, as well as terrestrial infrastructure based on mmWave access points (APs). We characterize the impact of UAV-BS backhaul dynamics on the system performance by comparing with the case of ideal backhauling. On top of this, we provide an updated review of 3GPP activities behind UAVs, IAB design, and NR-based relaying.



\section{Technology Background}\label{sect:02}

In this section, we outline the ongoing and planned 3GPP activities instrumental to employing UAVs for IAB in 5G NR systems and beyond. First, we concentrate on the UAV support in cellular systems: we overview use cases and challenges of UAV communications. Then, we proceed by introducing feasible IAB architectures and implementation options. Finally, we review the capabilities of underlying NR relays as an important technology enabler.

\subsection{UAV Support}

Over the recent years, UAV support and integration into the contemporary wireless systems has received significant industrial interest. Starting from Rel. 15, 3GPP has incorporated the corresponding capabilities into cellular standardization. In this context, the prospective Rel. 16 TR 22.829 summarizes the use cases and analyzes the UAV features that may require enhanced support. This includes live video broadcasting applications, command and control communications, and the use of UAV-BSs. The latter is specified in TR 38.811.


In Rel. 15 TR 36.777, 3GPP conducted a study on extended LTE support for aerial vehicles, which facilitates the use of cellular technologies by UAV-UEs. Initiated in 2017, this study summarizes possible cellular system improvements for efficient service of UAV traffic and its effects on the network. Particularly, it evaluates the performance of UAVs in urban and rural micro- and macrocell environments. Extensive simulations supplemented with field measurement data demonstrate that the usage of UAVs leads to increased uplink (UL) and downlink (DL) interference. Further, this work specifies important interference mitigation techniques. Another issue pertaining to high-density environments is UAV mobility. TR 36.777 identifies methods to provide additional path information that may be used in making mobility-related decisions.


Further areas that require development include aerial UE detection and identification~\cite{solomitckii2018technologies}. This relates to e.g., determining whether the UAV is permitted to fly. The 3GPP Rel. 16 TR 22.825 outlines the requirements for remote identification and tracking of UAVs linked to a cellular subscription. It also discusses the mechanisms for remote identification of UAVs. 

Currently, 3GPP continues to explore the ways for cellular systems to further support UAVs. This involves work on improving the mobility performance, business, security, and public safety needs for the purposes of identification. To that end, Rel. 16 TR 22.125 identifies the operating requirements for 3GPP systems. In this direction, 3GPP is expected to enhance the support of UAV connectivity and tracking in TR 23.754 and TR 23.755, which identify the applications that benefit from UAVs, and the corresponding architectural solutions. Clearly, UAVs are capable of accommodating a wide range of use cases for the emerging NR technology. One of such important scenarios is IAB.

\subsection{IAB Technology}

The standardization of the IAB was initially proposed by AT\&T, Qualcomm, and Samsung in the dedicated work item description, RP-171880. The corresponding study item led to the composition of TR 38.874, which summarizes all the activities related to the NR IAB.

Compared to terrestrial NR deployments, a major limitation of mmWave-based UAV-BSs is their backhaul link. Ground APs typically have a fixed wired backhaul connection and can offer very high data rates to the core network, whereas UAV-BSs should rely solely on wireless backhauling. With the introduction of NR systems, which support highly directional antenna arrays, UAV-BSs equipped with the IAB functions (named here UAV-based IAB) may facilitate on-demand network densification, thus efficiently avoiding interference and reducing capital investments into mobile infrastructure.

3GPP does not enforce any particular IAB implementation, which leaves specific details for vendors to decide upon. From the radio network planning perspective, available options include single-hop vs. multi-hop implementations, in-band vs. out-of-band backhauling, various radio technologies in use by the access link, as well as the levels of access and backhaul integration.

 
Although multi-hop backhauling can offer a performance boost for the network coverage known as range extension, it also brings additional overheads in terms of signaling. On top of conventional network management procedures, both multi-hop and single-hop IAB systems need to enable relay-specific functionality, such as backhaul link discovery and management; backhaul/access resource allocation and coordination; backhaul cross-link interference management, etc. Finally, there is a number of 3GPP-specific protocol-related options considered in TR 38.874, which account for the ways to realize multi-hop forwarding for both the UE and the relay nodes.


The radio technology and its frequency band~\cite{mourad2016self} on the access and backhaul links is another system design choice. For example, if both connections are implemented with the same radio, there is a possibility to utilize joint resource allocation mechanisms, which may reduce the overall system capacity but can also lower the deployment costs. Further, if both access and backhaul links operate over the same frequency band, there is a need for additional interference management. According to TR 38.874, the IAB-node should be capable of providing multi-radio access functionality, with at least Rel. 15 NR and legacy LTE options.


On top of these architectural choices, the implementation of IAB may differ with respect to the levels of backhaul and access integration. One can implement separate PHY and MAC realizations for access and backhaul, while sharing certain elements of MAC scheduling in a common module~\cite{polese2018end}. The crucial enabling technology for the IAB -- NR-based relying -- is currently being ratified by 3GPP.

\subsection{NR Relaying}


NR relaying with IAB capabilities has been discussed in TR 38.874 -- initially planned for Rel. 15 and now continued with the focus on Rel. 16. The role of the relay systems is to connect the UE with the donor BS, which is directly anchored on the core transport network. The gaps in NR coverage due to non-line-of-sight (NLOS) or blockage conditions remain a fundamental limitation for prospective NR deployments. In this context, the benefits of IAB relaying for NR are related to densification of the access network for increased reliability without the need to densify the associated transport network.


Accounting for the natural traffic aggregation at the backhaul links, the backhauling of traffic from the relay to the donor BS is attempted over the NR links. Similarly to the NR BS, the NR relay node may operate in stand-alone or non-stand-alone regimes as described in TR 38.912, such that most of the technology for NR access in Rel. 15 (see TS 38.300) is reused for backhaul links.


The current evaluation results in 3GPP contributions demonstrate considerable benefits from the use of single-hop relaying in the form of decreased outage and increased UE throughput as summarized in TR 38.874. To further exploit the relaying benefits, multi-hop support is presently being discussed by 3GPP. The performance evaluation studies (see R1-1812199, R1-1812982, and R1-1813418) confirmed the benefits of multi-hop relaying over the single-hop case for the users with poor channel conditions. The outlined reports contain a comprehensive account on the effects of the number of hops and load balancing considerations for the final metrics of interest. The anticipated extra gains from having additional hops, however, pose challenges related to selecting the best route, optimizing resource allocation, etc. Despite these, flexible multi-hop relaying topology is considered as one of the key components in future B5G systems to connect the UE with the core network. Also, TR 38.874 focuses on IAB with physically fixed relays. Importantly, it does not preclude from optimization for mobile relays in future releases.

The forthcoming Rel. 17 is expected to continue the studies of NR relaying across multiple use cases. One of the considered scenarios discussed in TR 22.866 is the relay support for enhanced energy efficiency and coverage. For example, an industrial factory may rely upon multiple UEs acting as relays to forward traffic from the target UE to its serving BS.


\section{Capturing Features of IAB Deployments}\label{sect:03}


A broader challenge of radio network optimization has been discussed thoroughly in the past few years. A number of efficient solutions have been proposed for various system settings in heterogeneous networks. However, these do not incorporate the unconstrained mobility of UAV-BSs as well as intricate features of the IAB design, such as the need to account for the NR propagation effects, unconstrained 3D deployment of UAVs, dynamic UAV-BS associations, and bandwidth partitioning between the AP and the UAV-BS, among other factors.  


Recently, the authors in \cite{kovalchukov2017modeling} demonstrated that capturing unconstrained 3D deployment of communicating entities inherent for IAB is critical for NR systems operating with directional antennas. However, higher altitudes of UAVs may partially alleviate the problem of dynamic human blockage. At the same time, adding the third dimension is also known to alter the mmWave propagation specifics. Further, it is crucial to account for the realistic user placement and mobility patterns.


IAB systems providing coverage extension and capacity boost are expected to be deployed on top of the terrestrial NR infrastructure. Serving the moving users, UAVs may continuously adjust their positions and thus the backhaul association point to the anchor BS. Similarly, in realistic deployments, users should also be able to dynamically change their network association point. As a result, a suitable performance optimization algorithm has to be flexible enough to capture these aspects while providing optimized performance at all times.


In the past works, the authors assumed static deployments of UEs across a certain area of interest when addressing the positioning of UAV-BSs in 3D space, such that a certain parameter of interest is optimized. However, these studies do not generally account for the aforementioned specifics of IAB implementation. Recent studies advocate for the use of dynamic optimization methodologies for IAB design \cite{tafintsev2018improved}. In contrast to conventional techniques that assume static UEs deployments, new adaptive algorithms need to accommodate changes in UE locations by updating the UAV-BS placement within a bounded 3D space to fully benefit from their inherently mobile nature.


In summary, the latest work indicates that accurate performance assessment and optimization of the IAB-based NR design requires a number of modeling choices to be specified carefully. Particularly, one needs to (i) account for true 3D layouts for UAV-BS positioning, (ii) rely upon accurate air-to-ground NR propagation models, (iii) employ realistic UE deployment and mobility models, and (iv) incorporate practical deployment considerations. In the next section, we demonstrate how these angles come together in applying optimization algorithms to efficiently address the problem of mmWave-based IAB design.

\section{Performance of UAV-based IAB Systems}\label{sect:04}

In this section, we explore the performance of UAV-based IAB systems in urban deployments with clustered mobile users. Particularly, we demonstrate the gains of utilizing optimized UAV-based IAB operation over the conventional grid deployment as well as characterize the trade-offs arising from converged aerial and terrestrial communications.

\subsection{Deployment Considerations}

\begin{figure}[!t]
    \centering
    \includegraphics[width=\columnwidth]{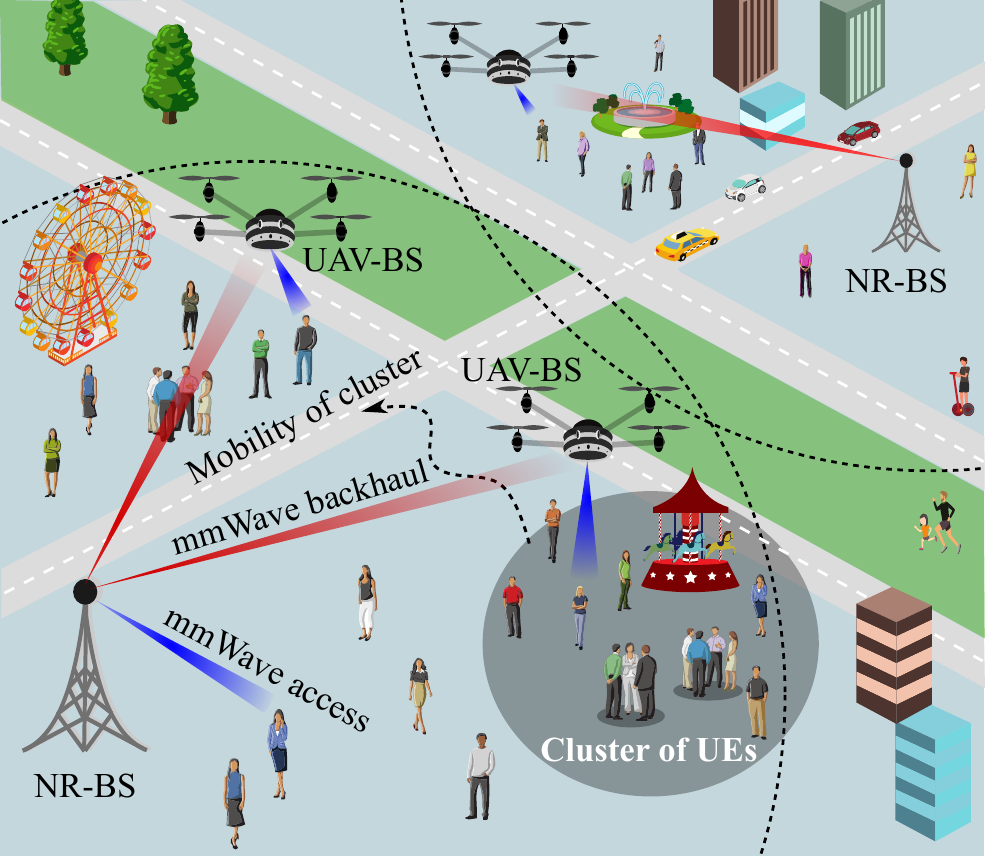}
    \caption{Usage of mmWave-based UAV-BSs.}
    \label{fig:basics}
\vspace{-3mm}
\end{figure}


We consider a square area covered by terrestrial NR APs and aerial UAV-BSs offering additional connectivity options for the UEs. We concentrate on studying a relay topology, where APs have two interfaces for the UEs and UAV-BSs (see Fig.~\ref{fig:basics}). Further, we assume that based on the signal strength the UE may connect either to the UAV-BS or to the ground AP. The altitude and the speed of UAV-BSs are fixed, and the latter can only alter their directions in the horizontal plane. We concentrate on the UL channel from UEs to APs by considering constant traffic from the UEs.

In the addressed scenario, APs act as traffic sinks, since the UL traffic from the UEs to UAV-BSs is further forwarded to the APs. We also require packet buffers on the UAV-BSs and UEs. Packets are queued at the UAV-BSs: if the UAV-BS buffer is full, it drops any arriving packets. In this work, we assume separate channels for access and backhaul links as well as dedicated antenna arrays for each interface. In practice, it means that an IAB node is equipped with two separate PHY interfaces, which run independent MAC schedulers, whereas routing, packet queuing, and other procedures are coordinated by the common IAB entity.

\begin{figure}[!t]
    \centering
    \includegraphics[width=\columnwidth]{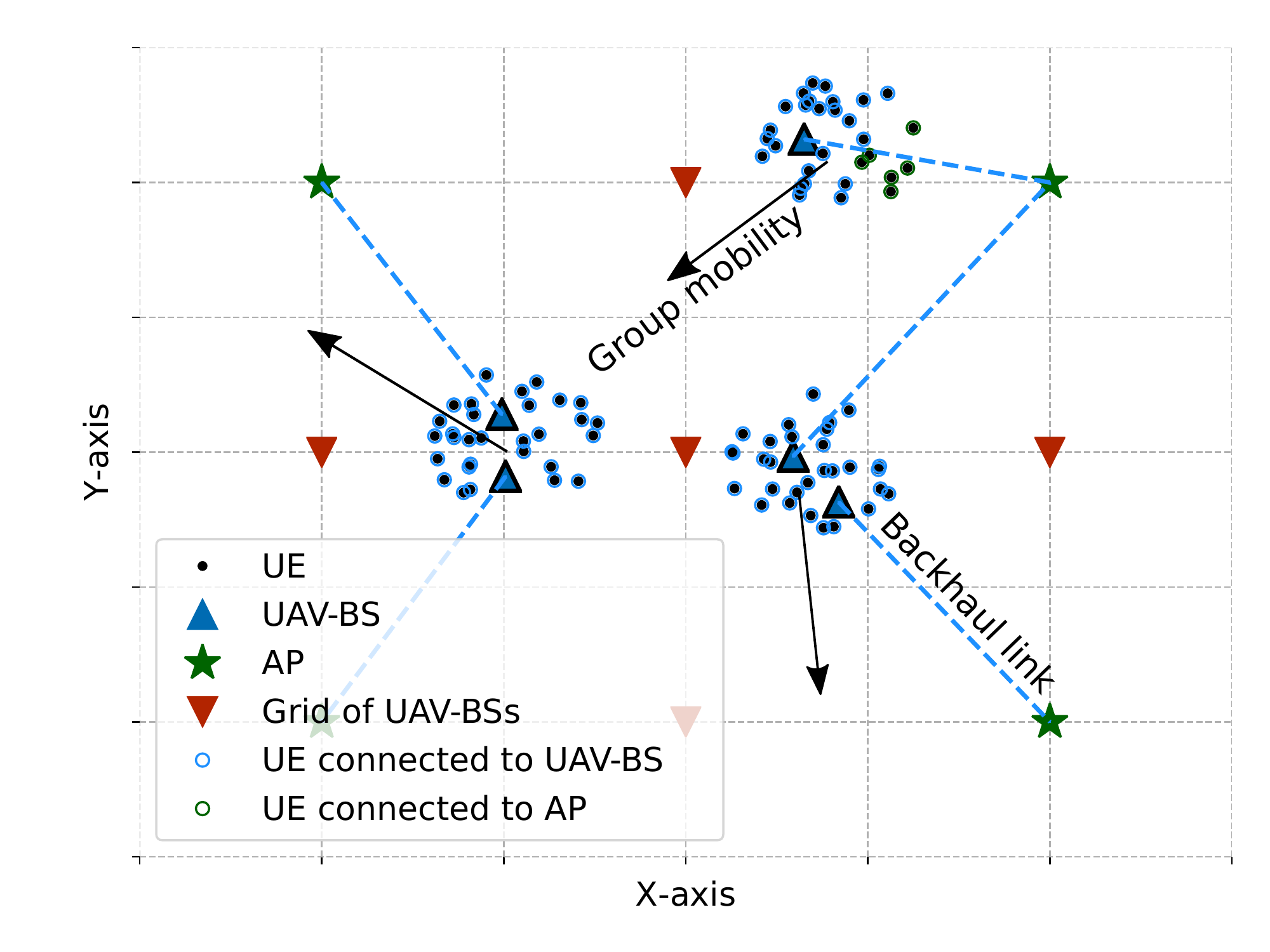}
    \caption{Example of PSO-based and grid deployment of UAV-BSs.}
    \label{fig:deployment}
\vspace{-3mm}
\end{figure}

In what follows, we investigate scenarios where the spatial density of users varies over time (see Fig.~\ref{fig:deployment}). To model user mobility, we employ the Reference Point Group Mobility (RPGM) model~\cite{bai2004survey}. Accordingly, each group of users has a leader, whose movement determines the mobility direction of the entire group. At the UE to AP and UE to UAV-BS interfaces, we follow $\alpha$-fairness as a broad class of utility functions that capture different fairness criteria~\cite{gerasimenko2015cooperative}. The default system parameters are summarized in Table~\ref{table:parameters}.

\begin{table}[!b]
\vspace{-3mm}
\renewcommand{\arraystretch}{1}
\caption{Assumed System and Evaluation Parameters.}
\label{table:parameters}
\centering
\begin{tabular}{p{0.45\columnwidth}|p{0.45\columnwidth}}
\hline
\hline
\bfseries Parameter & \bfseries Value \\
\hline
Deployment area & 600 m x 600 m\\
\hline
Number of APs & 4\\
\hline
Number of UAV-BSs & 2\\
\hline
Number of UEs & 100\\
\hline
Distribution of UEs & Clustered\\
\hline
Radius of UE distribution & 50 m\\
\hline
Seeds & 200\\
\hline
Channel model for AP-UAV link & mmWave Channel Model LOS\\
\hline
Channel model for AP-UE link & mmWave Channel Model\\
\hline
Channel model for UAV-UE link & mmWave Channel Model\\
\hline
AP planar antenna arrays & 8x8 (access and backhaul)\\
\hline
UAV-BS planar antenna arrays & 8x8 (access), 4x4 (backhaul)\\
\hline
UAV-BS altitude & 20 m\\
\hline
AP height & 20 m\\
\hline
UE height & 1.5 m\\
\hline
Carrier frequency & 73 GHz\\
\hline
System bandwidth & 0.56 GHz\\
\hline
Transmit power &  24 dBm\\
\hline
Power control & Full-power \\
\hline
Beam-sweeping periodicity & 3 $\mu$s\\
\hline
Frame size & 3 $\mu$s\\
\hline
Transmission mode & TDM\\
\hline
Packet size & 1000 bytes\\
\hline
Target data rate & 500 Mb/s\\
\hline
AP and UAV-BS scheduling & Round-Robin \\
\hline
Value of $\alpha$ & 2 \\
\hline
\end{tabular}
\end{table}

\subsection{Approach and Metrics}


For the purposes of UAV-based IAB optimization, we consider a dedicated dynamic algorithm based on particle swarm optimization (PSO) method. PSO is a useful heuristics that addresses dynamic optimization by iteratively improving a solution with respect to a given parameter. This algorithm emulates the interactions between particles to share information. It solves the problem by having a set of possible solutions in the feasible region of a given problem. The movement of each particle is influenced by its local best-known position but is also guided toward the best-known positions in the search space, which are updated as better positions are being discovered by other particles.


The numerical assessment is conducted with our custom-made system-level simulator named WINTERsim, which has been utilized extensively for 5G/5G+ performance evaluation~\cite{gerasimenko2015cooperative}. This simulation environment was further extended to support UAV-BSs in accordance with the recent 3GPP requirements on aerial access. We consider two metrics of interest: the mean UE throughput and fairness. For the latter, we use the Jain's fairness index, which quantifies the ''equality'' of UE performance. If all the UEs receive the same throughput, this index equals $1$. Generally, it ranges from $1$ to $1/N$, where $N$ is the number of UEs. The analytical framework for the optimal trade-off between performance and fairness is elaborated in \cite{7876861} and \cite{6547814}. These works establish an approach for maximizing the performance of a wireless system for a given fairness.

\subsection{Numerical Results and Assessment}


\begin{figure}[!t]
    \centering
    \includegraphics[width=0.9\columnwidth]{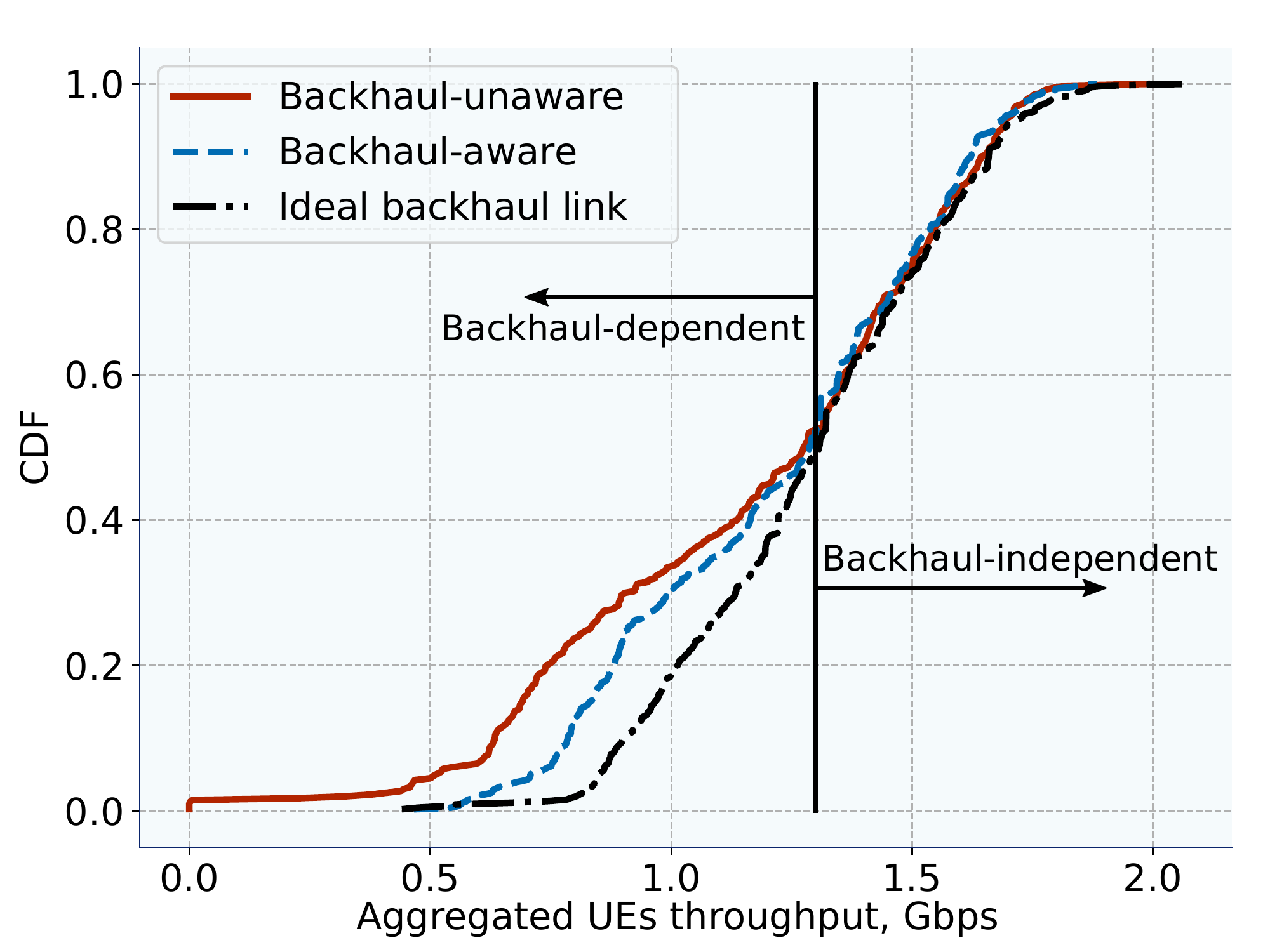}
    \caption{CDF of aggregated UEs throughput.}
    \label{fig:thr_cdf}
\end{figure}

We proceed with providing the representative numerical results for UAV-based IAB operation. We start by highlighting the importance of accounting for constraints imposed by backhauling in the considered integrated terrestrial/aerial deployment. Fig.~\ref{fig:thr_cdf} reports on the cumulative distribution function (CDF) of the mean UE throughput for the considered dynamic PSO optimization algorithm in three different cases: (i) ideal backhaul, (ii) backhaul-unaware, and (iii) backhaul-aware. The former assumes that the backhaul bandwidth is unlimited and may potentially accommodate all the requirements imposed by the UEs at the UE-UAV interface. In the backhaul-unaware scheme, we enforce a backhaul bandwidth rate limitation but it is not accounted for by the optimization algorithm. Finally, backhaul-aware scheme explicitly incorporates the rate limitation at the backhaul air interface.


Analyzing the results in Fig.~\ref{fig:thr_cdf}, one may observe that the ideal backhaul scheme significantly overestimates the actual throughput, since the probability of having the throughput of less than about 0.7 Gbps is negligibly small. This is explained by the fact that all of the traffic generated by the UEs in the UAV-BS coverage area is assumed to be delivered to the APs successfully. This assumption allows the optimization algorithm to position the UAV-BSs directly on top of the user clusters, thus maximizing the throughput at the UE-UAV air interface. Conversely, the backhaul-unaware scheme drastically underestimates the actual throughput, as it does not explicitly account for the backhaul rate limitation. In this case, the UAV-BSs are positioned similarly to those in the ideal backhaul case but the rate at the UAV-AP interface is often insufficient to offload all of the generated traffic.

Another interesting observation in Fig.~\ref{fig:thr_cdf} is that the performance of all the considered schemes differs only for low-to-medium throughput allocations, which means that starting from approximately 1.3 Gbps the curves coincide with each other. This is a consequence of having the throughput limitations on the backhaul links. In particular, the throughput of the best-located AP is the same in all of the cases, which yields that for the cluster of UEs deployed close to the AP the backhaul link does not have much impact.

\begin{figure}[!t]
    \centering
    \includegraphics[width=0.9\columnwidth]{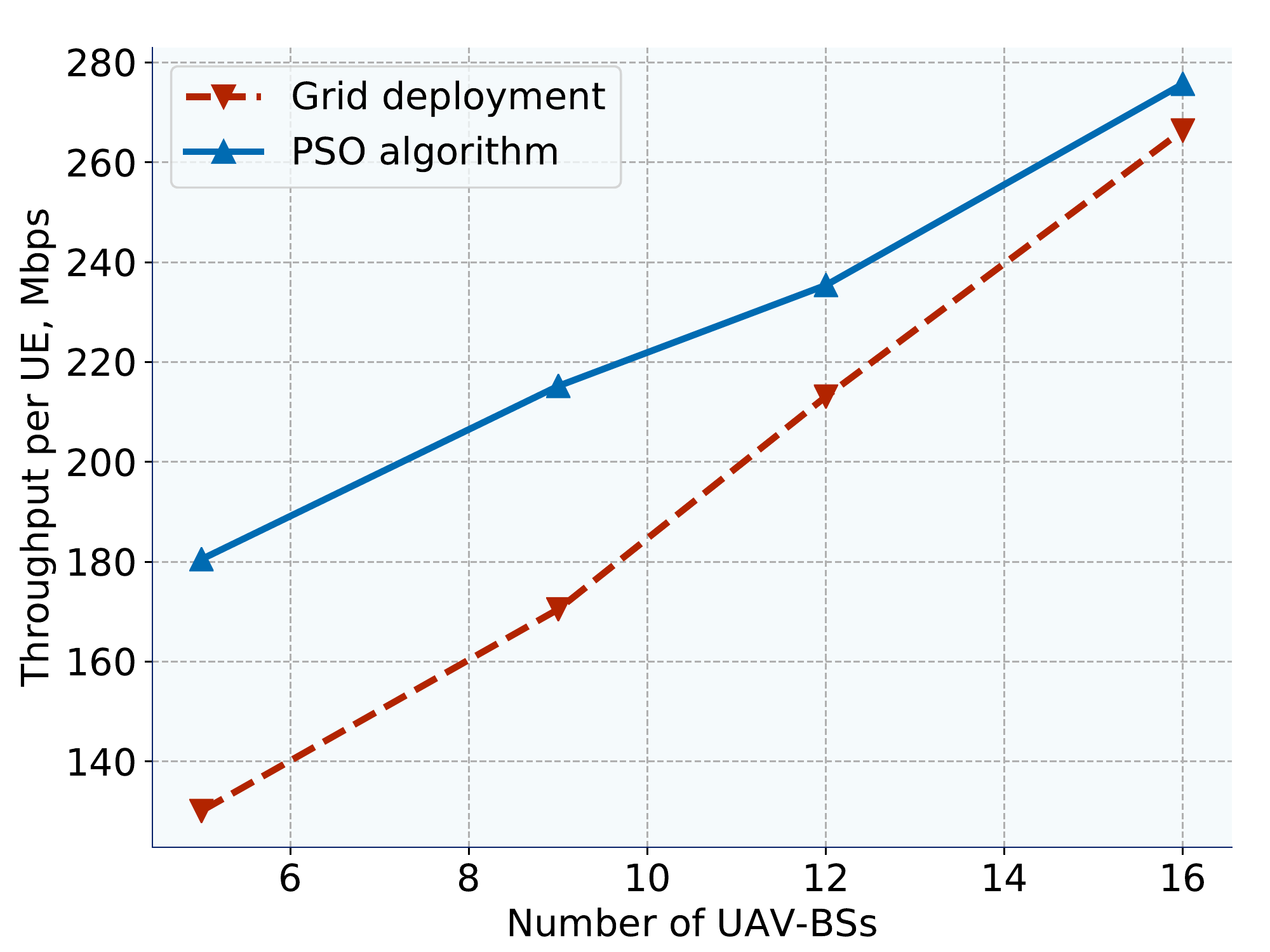}
    \caption{Mean UE throughput for different numbers of UAV-BSs.}
    \label{fig:thr_drones}
\end{figure}

It is natural to expect that the UAV-BS positions and thus the UE throughput allocations will heavily depend on the interplay between the number of clusters and the number of UAV-BSs. We study this effect by using the mean UE throughput as our parameter of interest. In Fig.~\ref{fig:thr_drones}, we vary the number of UAV-BSs deployed in the scenario while keeping the number of clusters equal to $4$. This graph also compares the results for the dynamic PSO-based algorithm with those for the static grid-based deployment of UAV-BSs. As expected, for both deployments the mean UE throughput increases as the number of UAV-BSs grows. The most significant difference between the deployments is observed for a small number of UAV-BSs.

The rationale behind this behavior is that here the grid-based deployment cannot fully serve the area of interest, and occasionally the UEs either end up far away from the UAV-BSs or may not be under their coverage. As the number of nodes increases, this difference decreases, since the grid-based deployment now densely spans across the entire service area and all of the UEs are always within coverage of at least one UAV-BS. Another reason for the reduced gains is that with an extremely high number of UAV-BSs the interference starts to play an important role for the achievable throughput by not allowing the algorithm to position the UAV-BSs as close as needed.

\begin{figure}[!b]
    \centering
    \includegraphics[width=0.9\columnwidth]{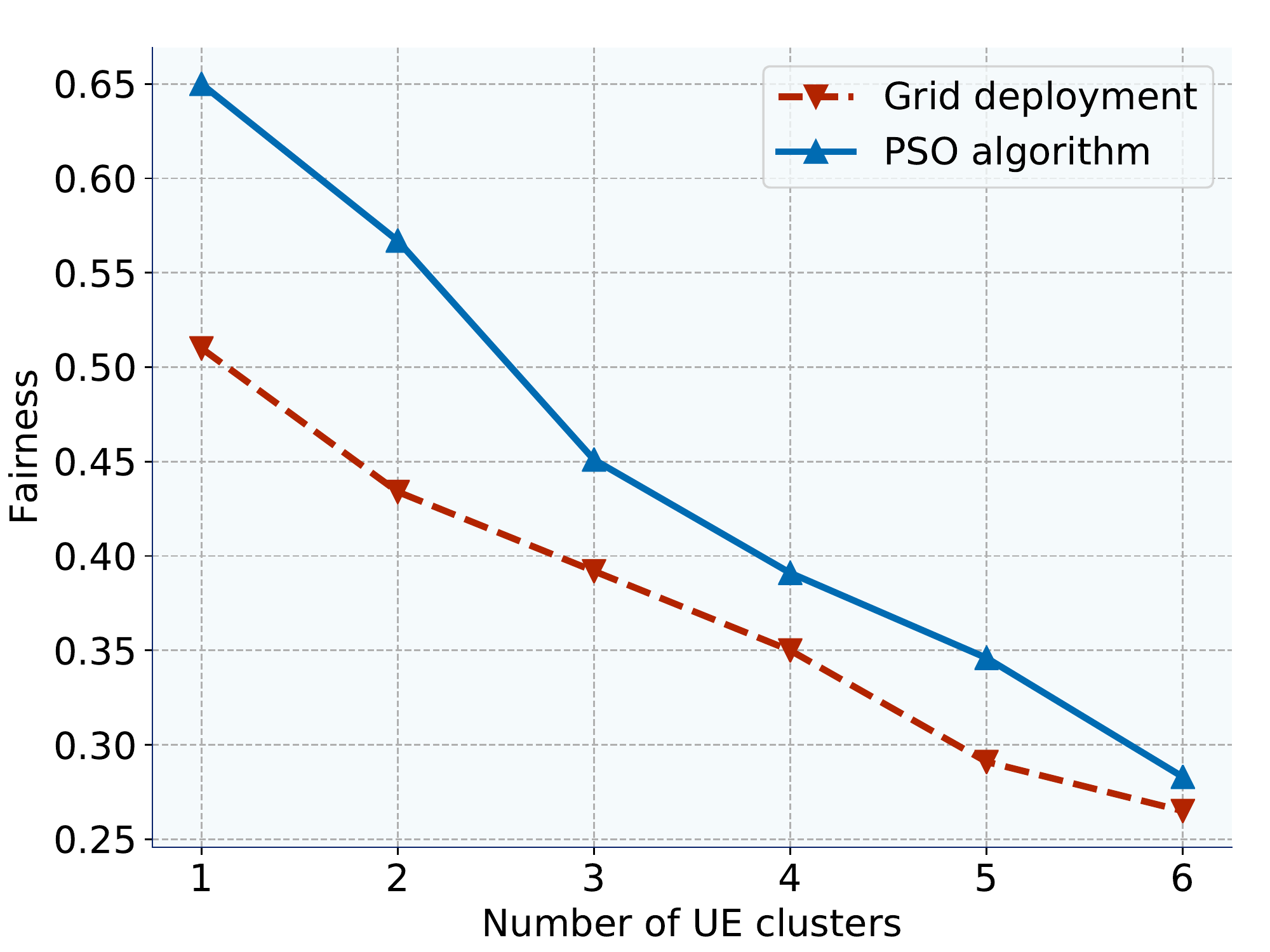}
    \caption{Fairness for different numbers of UE clusters.}
    \label{fig:jain_index}
\end{figure}


The considered dynamic PSO algorithm is useful for determining the optimized UAV-BS locations to maximize the UE throughput based on the $\alpha$-fairness criterion. In Fig.~\ref{fig:jain_index}, we study the impact of the number of clusters on the Jain's fairness index by keeping the number of UAV-BSs constant at $5$ for the two considered UAV-BS deployments. First, it is important to note that an increase in the number of UE clusters leads to a drop of fairness between the throughput allocations. Particularly, there is a significant degradation for three clusters. The underlying reason for this in case of dynamic PSO optimization is that for one, two, or four clusters there are distinct ``optimal'' combinations of the UAV-BS positions, which lead to similar UE performance. As the number of clusters increases, there are no such positions, which results in equal distances to the associated UAV-BSs. For the grid-based deployment, one may observe a similar trend -- the fairness of UE throughput allocations decreases.

Similarly to Fig.~\ref{fig:thr_drones}, the highest gains in fairness are achieved for a small number of UE clusters. Already for three clusters, the fairness values for the two deployment schemes become close to each other. Even though the PSO-based solution still demonstrates better performance, the UAV-BS grid is composed in a way that maximizes the probability for the UEs to be covered by at least one UAV-BS or a ground AP. The performance levels of the two considered deployments become ever more similar when the number of clusters is higher than the number of UAV-BSs.

\section{Conclusions}\label{sect:05}

Integrated Access and Backhaul is an emerging technology considered by 3GPP for cellular densification and coverage extension. Future UAV-based IAB systems operating in mmWave spectrum therefore become a promising solution due to their higher data rates at the access and backhaul links, lower interference, and more flexible service dynamics. However, the deployment of such systems is subject to multiple restrictions related to the infrastructure network support, resource partitioning between the access and the backhaul links, as well as correlated mobility of the UEs and UAV-BSs.

In this work, we provided a detailed account of the recent 3GPP activities behind UAV-based IAB systems that operate in mmWave frequencies. In an environment with realistic UE mobility across the terrestrial NR layout, we characterized the important factors and trade-offs related to UAV-aided IAB operation. Specifically, we demonstrated that dynamic methodologies that optimize the UAV-BS positions can unlock substantial gains as compared to rigid deployments. Our numerical results show the gains of approximately 0.2 Gbps for the backhaul-aware solution as well as demonstrate that optimization improves the overall system performance. In this context, the constraints imposed by mmWave backhauling become crucial for the effective performance of UAV-based IAB systems.


\bibliographystyle{IEEEtran}
\bibliography{bibliography}


\textbf{Nikita Tafintsev} is currently a Research Assistant with the Unit of Electrical Engineering, Tampere University, Finland. He received B.Sc. degree with honors in Radio Engineering, Electronics and Telecommunication Systems from Peter The Great St. Petersburg Polytechnic University, St. Petersburg, Russia, in 2017. His research interests include performance evaluation and optimization methods for mmWave networks, UAV communications, and 5G systems.

\textbf{Dmitri Moltchanov} received M.Sc. and Cand.Sc. degrees from the St. Petersburg State University of Telecommunications, in 2000 and 2003, respectively, and the Ph.D. degree from the Tampere University of Technology in 2006. Currently, he is University Lecturer with the Unit of Electrical Engineering, Tampere University, Finland. In his career, he has taught more than 50 courses on wireless and wired networking and (co-)authored more than 150 papers. His research interests include 5G/5G+, New Radio, industrial IoT, and V2V/V2X.

\textbf {Mikhail Gerasimenko} is a post-doctoral Researcher at Tampere University, Finland. He obtained M.Sc. in 2013 and Ph.D. degree in 2018 from Tampere University of Technology. Mikhail started his academic career in 2011 and since then he appeared as (co-)author of multiple scientific publications. His main subjects of interest are machine-type communications and heterogeneous networks. 

\textbf{Margarita Gapeyenko} is a Ph.D. candidate at the Unit of Electrical Engineering at Tampere University, Finland. She earned M.Sc. degree in Telecommunication Engineering from University of Vaasa, Finland, in 2014, and B.Sc. degree in Radio Engineering, Electronics, and Telecommunications from KSTU, Kazakhstan, in 2012. Her research interests include mathematical analysis, performance evaluation, and optimization methods for mmWave networks, UAV communications, and (beyond-)5G heterogeneous systems.


\textbf{Jing Zhu} received B.S. and M.S. degrees from Tsinghua University, China, in 1999 and 2001, respectively, and a Ph.D. in 2004 from the University of Washington, Seattle, all in electrical engineering. He is currently a principal engineer at Intel Corporation. His main research interests are system design, performance optimization, and applications for heterogeneous wireless networks, including 4G/5G cellular systems, high-density wireless LANs, and mobile ad hoc networks.

\textbf{Shu-ping Yeh} received the B.S. degree from National Taiwan University in 2003, and the M.S. and Ph.D. degrees from Stanford University in 2005 and 2010, respectively, all in electrical engineering. She is currently a Research Scientist with Intel Labs. Her recent research focus includes advanced self-interference cancellation technology, full-duplex PHY/MAC system designs, and multi-tier multi-RAT heterogeneous networks.

\textbf{Nageen Himayat} is a Principal Engineer with Intel Labs, where she leads a team conducting research on several aspects of next generation (5G/5G+) of mobile broadband systems. Dr. Himayat obtained her B.S.E.E degree from Rice University, and her Ph.D. degree from the University of Pennsylvania. Her research contributions span areas such as multi-radio heterogeneous networks, mmWave communication, energy-efficient designs, cross layer radio resource management, multi-antenna, and non-linear signal processing techniques.

\textbf{Sergey Andreev} is an assistant professor of communications engineering and Academy Research Fellow at Tampere University, Finland. He received his Ph.D. (2012) from TUT as well as his Specialist (2006) and Cand.Sc. (2009) degrees from SUAI. He (co-)authored more than 200 published research works on intelligent IoT, mobile communications, and heterogeneous networking.

\textbf{Yevgeni Koucheryavy} received the Ph.D. degree from the Tampere University of Technology, in 2004. He is a Professor with the Unit of Electrical Engineering, Tampere University, Finland. His current research interests include various aspects in heterogeneous wireless communication networks and systems, the IoT and its standardization, and nanocommunications.

\textbf{Mikko Valkama} received his M.Sc. and Ph.D. degrees (both with honors) from Tampere University of Technology in 2000 and 2001, respectively. Currently, he is a Full Professor and Head of Unit of Electrical Engineering at Tampere University (TAU), Finland. His general research interests include radio communications, radio localization, and radio-based sensing, with particular emphasis on 5G and beyond mobile radio networks.

\end{document}